\documentclass[aps,pre,twocolumn,superscriptaddress,showpacs]{revtex4}

\usepackage{graphicx}
\usepackage{dcolumn}
\usepackage{bm}
\usepackage{color}

\begin{document}


\title{Bubble statistics and positioning in superhelically stressed DNA}

\author{Daniel Jost}
\affiliation{Laboratoire de Physique and Centre Blaise Pascal of the \'Ecole Normale Sup\'erieure de Lyon, Universit\'e de Lyon, CNRS UMR 5672, Lyon, France}
\affiliation{Department of Physics, Harvard University, Cambridge, MA 02138, USA}

\author{Asif Zubair}
\affiliation{Laboratoire de Physique  and Centre Blaise Pascal of the \'Ecole Normale Sup\'erieure de Lyon, Universit\'e de Lyon, CNRS UMR 5672, Lyon, France}

\author{Ralf Everaers}
\affiliation{Laboratoire de Physique  and Centre Blaise Pascal of the \'Ecole Normale Sup\'erieure de Lyon, Universit\'e de Lyon, CNRS UMR 5672, Lyon, France}

\date{\today}

\begin{abstract}

We present a general framework to study the thermodynamic denaturation of double-stranded DNA under superhelical stress.
We report calculations of position- and size-dependent opening probabilities for bubbles along the sequence. Our results are obtained from transfer-matrix solutions of  the Zimm-Bragg model for unconstrained
DNA and of a self-consistent linearization of the Benham model for superhelical DNA.
The numerical efficiency of our method allows for the analysis of entire genomes and of random sequences of
corresponding length ($10^6–10^9$ base pairs).
We show that, at physiological conditions, opening in superhelical DNA is strongly
cooperative with average bubble sizes of $10^2–10^3$ base pairs (bp), and orders of magnitude higher than in unconstrained DNA. In heterogeneous sequences, the average degree of base-pair opening is self-averaging, while bubble localization and statistics are dominated by sequence disorder. Compared to random sequences with identical GC-content, genomic DNA has a significantly increased probability to open large bubbles under superhelical stress. These bubbles are frequently located directly upstream of transcription start sites. 
\end{abstract}

\pacs{87.14.gk, 87.15.A-, 36.20.Ey,}

\maketitle

\section{Introduction}

Fundamental processes, such as transcription and replication require a transient, local opening of the DNA double helix \cite{calladine,alberts}. Such {\it bubbles} occur spontaneously under physiological conditions \cite{kowalski88,kowalski89}, while complete melting and the separation of the two complementary strands requires temperatures around $80^o$C \cite{FK}. Bubbles have been implicated as an explanation for high cyclization rates in short DNA fragments \cite{cloutier,yan}, but their main interest lies in biology and in the physical mechanism underlying the functioning and control of transcription and replication start sites: the stability of DNA is sequence-dependent \cite{SL} and opening is strongly influenced by superhelical stress \cite{benham92,fye99,benham06} and the binding of regulatory proteins \cite{sobellPNAS85,ambjornsson05}. In particular, Benham and coworkers showed significant correlations between the positions of strongly stress-induced destabilized regions and regulatory sites \cite{wang04,ak05,wang08}. \\
Several models exist to describe the internal opening of DNA. 
The Peyrard-Bishop-Dauxois \cite{peyrard89,dauxois93} and the Poland-Scheraga (PS) models \cite{poland,jostBJ09} have already been used to quantify bubble statistics \cite{kalosakas04,vanerp05,ares07,kafri02,blossey03,garel05,coluzzi07,monthus07} and for the {\it ab initio} annotation of genomes on the basis of correlations between biological function and thermal melting \cite{yeramian02,carlon07,jostJPCM09}. 
Here, we describe the bubble statistics of random and biological sequences at physiological conditions using (i) the Zimm-Bragg (ZB) model, an efficient approximation of the PS model \cite{jostJPCM09}, and (ii) the Benham model \cite{fye99}, a generalization of the ZB model accounting for superhelical density but neglecting writhe. 
After exploring the bubble statistics of unconstrained DNA using the ZB model, we introduce an asymptotically exact self-consistent linearization \cite{vologodskii} of the Benham model as a precise and convenient tool to study the huge impact of superhelicity on local bubble opening.  The numerical efficiency of the method allows us to investigate the bubble statistics for entire genomes and random sequences of sufficient length ($10^6-10^9$ bp) to obtain statistically significant results for sequence effects on bubble statistics and positioning. 
In a final step, we correlate the positions of highly probable bubbles within the genome of {\it E. coli} with the position of transcription start sites (TSS) and start codons.

\section{Models and Methods}
 
\subsection{Unconstrained DNA}

\subsubsection{The Zimm-Bragg model}

The most widely-used model to treat the denaturation of DNA chains is the PS model \cite{poland} which offers predictive power for thermal melting and strand dissociation for DNA of arbitrary length, strand concentration and a wide range of ionic conditions.
For long heterogeneous sequences, whose local denaturation is dominated by the quenched sequence disorder \cite{garel05,coluzzi07,monthus07}, the related and computationally faster ZB model gives surprisingly good results \cite{jostJPCM09}.
In the PS and ZB formalisms, the free energy of a given configuration is decomposed into formation free energies for closed base-pair steps and free energy penalties for the nucleation of unpaired regions (or bubbles). Using periodic boundary conditions, the ZB Hamiltonian for a circular chain of length $N$ can be expressed as \cite{jostJPCM09}
\begin{equation}
\mathcal{H} _{ZB} = \sum_{i=1}^N \left\{ (\Delta g_{i}-\Delta g_{loop})\theta_i\theta_{i+1}+\Delta g_{loop}\theta_i\right\}  \label{eq:zb}
\end{equation}
where $\theta_i=0$ ($1$) if base-pair $i$ is open (closed). $\Delta g_{i}$ ($\sim k_b T $)  is the nearest-neighbor (NN) free energy to form the base-pair step $(i,i+1)$.
$\Delta g_{loop}\equiv-k_B T\log[\sigma_{PS}\bar{L}^{-c}]$ is the loop nucleation penalty and depends on the PS cooperativity $\sigma_{PS}\sim 10^{-4}$, on the interacting self-avoiding loop exponent $c\sim2.1$ \cite{kafri02,blossey03} and on a typical bubble length $\bar{L}$ \cite{jostJPCM09} ($\Delta g_{loop}\sim 20 k_b T$ for $\bar{L}=170$). Parameters are taken at physiological conditions for temperature $T=37^o$ C and salt concentration $[Na^+]=0.1$ M \cite{jostJPCM09}.  

\subsubsection{Transfer-matrix method}

Being formulated as a 1D Ising model, the model is easily solved analytically (numerically) for homogeneous (heterogeneous) sequences using transfer matrix methods. The partition function of the system is then given by \cite{jostJPCM09}
\begin{equation}\label{eq:zbz}
\mathcal{Z}=\textrm{Trace}\left[\prod_{i=1}^{N} T_i\right],
\end{equation}
the individual closing probability by
\begin{equation}\label{eq:zbthet}
\langle \theta_i\rangle=\frac{1}{\mathcal{Z}}\textrm{Trace}\left[\left(\prod_{j=1}^{i-1} T_j \right) T'_i, \left(\prod_{j=i+1}^{N} T_j \right)\right].
\end{equation}
In particular, it is possible to calculate position resolved opening probabilities, $P_{i,l}$, for bubbles containing $l$ open bps and beginning at the closed bp $i$, by
\begin{equation}\label{eq:zbpb}
P_{i,l}=\frac{1}{\mathcal{Z}}\textrm{Trace}\left[\left(\prod_{j=1}^{i-1} T_j \right) T'_i S T'_{i+l+1} \left(\prod_{j=i+l+2}^{N} T_j \right)\right]
\end{equation}
with 
\begin{displaymath}
T_i = \left( \begin{array}{cc}
e^{-\beta \Delta g_{i}} & e^{-\beta\Delta g_{loop}} \\
1 & 1
\end{array} \right),
\end{displaymath}
\begin{displaymath}
T'_i = \left( \begin{array}{cc}
e^{-\beta \Delta g_{i}} & e^{-\beta\Delta g_{loop}} \\
0 & 0
\end{array} \right)\quad\textrm{and}\quad S = \left( \begin{array}{cc}
0 & 0 \\
1 & 1
\end{array} \right).
\end{displaymath}
The global closing probability $\Theta$ and the probability per bp $P_l$ to observe a bubble of length $l$ are given by $\Theta=(\sum_{i=1}^N \langle \theta_i\rangle)/N$ and $P_l=(\sum_{i=1}^N P_{i,l})/N$. Basic summing rules on the bubble probabilities imply that $1-\Theta = \sum_l l P_l$, and that $\mathcal{L}=(\sum_l l P_l)/(\sum P_l)=(1-\Theta)/P_b$,
\begin{equation}
\mathcal{L}=\frac{\sum_l l P_l}{\sum P_l}=\frac{1-\Theta}{P_b}
\end{equation}
where $\mathcal{L}$ is the average bubble length and $P_b=\sum_l P_l$ is the probability per bp to observe a bubble of arbitrary length.

For homogeneous sequences, these equations could be solved easily. In the asymptotic limit $N\rightarrow \infty$, it leads to
\begin{eqnarray}
\mathcal{Z} & = & \left[\frac{1+g+\sqrt{(g-1)+4s}}{2}\right]^N, \label{eq:homoz}\\
\langle\theta\rangle & = & \frac{1}{2}\left[1+\frac{g-1}{\sqrt{(g-1)^2+4s}}\right], \label{eq:homozb}\\
P_l & = &  \frac{1}{2}\left[\frac{2}{1+g+\sqrt{(g-1)^2+4s}}\right]^{l+2} \nonumber \\
& &\times\left[\frac{2s+g\left(g-1+\sqrt{(g-1)^2+4s}\right)}{\sqrt{(g-1)^2+4s}}\right] \label{eq:homop} 
\end{eqnarray}
with $g=\exp[-\beta\Delta g]$ and $s=\exp[-\beta\Delta g_{loop}]$.
The bubble probability is therefore a decreasing exponential function of $l$ for homogeneous sequences. 

For heterogeneous sequences, we numerically compute the different desired observables in a $\mathcal{O}(N)$-algorithm. In a first step, we iteratively compute the forward and backward products, $F_i=\left(\prod_{j=1}^{i} T_j \right)$ and $B_i=\left(\prod_{j=i}^{N} T_j \right)$. The second step consists in applying equations \ref{eq:zbz}, \ref{eq:zbthet} and \ref{eq:zbpb}.

\subsection{Superhelical DNA}

In living organisms, DNA is highly topologically constrained into circular domains (closed-loops or circular molecules)  \cite{calladine}.
Each closed domain is defined by a topological invariant  $L$, the so-called linking number. $L$ represents the algebraic number of turns either strand of the DNA makes around the other. It can be decomposed in two contributions: the twist which is the number of turns of the double-helix around its central axis, and the writhe which is the number of coils of the double-helix.
In the majority of living systems, the average linking number $L$ is below the characteristic linking number value $L_o$ of the corresponding unconstrained linear DNA, due to a negative superhelical density $\sigma=\alpha A/N\approx -0.06$ imposed by protein machineries, where $\alpha=L-L_o$ is the linking number difference.

\subsubsection{The Benham model}

\begin{figure}
\begin{center}
\includegraphics[width=8.5cm]{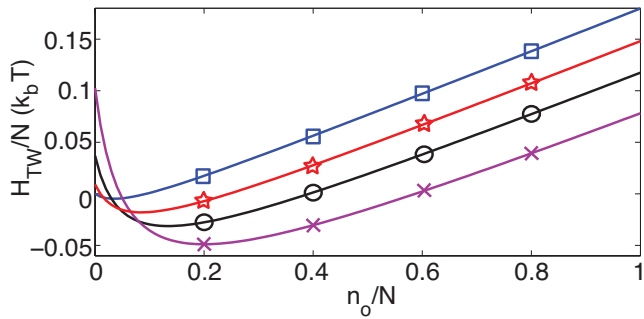}\caption{Effective superhelical hamiltonian $\mathcal{H}_{TW}$ per bp as a function of the relative opening $n_o/N$ in the asymptotic limit, for different stress values: $\sigma=0$ (squares), -0.03 (stars), -0.06 (circles) and -0.1 (crosses).}\label{fig:qeff}
\end{center}
\end{figure}
In this article, we consider bubble openings in superhelically constrained circular DNA using the Benham model for an imposed superhelical density $\sigma$, where the standard thermodynamic description of base-pairing ($\mathcal{H}_{ZB}$) is coupled with torsional stress energetics \cite{benham92,fye99}.
For each state, if one neglects the writhe contribution, the imposed linking difference $\alpha$ can be decomposed in three contributions: 1) the denaturation of $n_o$ base-pairs relaxes the helicity by $-n_o/A$ where $A=10.4$ bp/turn is the number of base-pair in a helical turn; 2) the resulting single-strand regions can twist around each other inducing a global over-twist of $\mathcal{T}$; 3) then, the bending and twisting of double-stranded parts is put in the residual linking number $\alpha_r$.  Therefore, due to the topological invariance of $\alpha$, we get the closure relation
\begin{equation}\label{eq:she}
\alpha=-\frac{n_o}{A}+\mathcal{T}+\alpha_r
\end{equation}  
For denatured regions, the high flexibility of single-stranded DNA allows unpaired strands to interwind. The energy associated with the helical twist $\tau_i$ (in rad/bp) of open base-pair $i$ is 
\begin{equation}
\mathcal{H}_{tw}(\theta_i,\tau_i)=\frac{C}{2} (1-\theta_i)\tau_i^2
\end{equation}
where the torsional stiffness $C$ is known from experiments, $C\approx 3.09 k_B T$.
The individual twist $\tau_i$ are related to the global over-twist $\mathcal{T}$ via the relation $\mathcal{T}=\sum_i (1-\theta_i)\tau_i/(2\pi)$. 
For paired helical regions, it has been experimentally found that superhelical deformations induce an elastic energy, quadratic in the residual linking difference
\begin{equation}
\mathcal{H}_r = \frac{K}{2}\alpha_r^2=\frac{K}{2}\left(\alpha+\frac{n_o}{A}-\mathcal{T}\right)^2
\end{equation}
where $n_o=\sum_i (1-\theta_i)$ is the number of open base-pairs and $K\equiv K'/N \approx 2220 k_B T /N$.
By integrating over the $n_o$ continuous degrees of freedom $\tau_i$, the superhelical stress energetics is represented by a non-linear effective Hamiltonian \cite{fye99}:
\begin{eqnarray}
\mathcal{H}_{TW}(n_o)  & = &\frac{2\pi^2 C K}{4\pi^2C+Kn_o}\left[\alpha+\frac{n_o}{A}\right]^2 \nonumber \\
  & &-\frac{1}{2\beta} \log\left(\left[\frac{2\pi}{ \beta C}\right]^{n_o}\frac{4\pi^2 C}{4\pi^2 C+K n_o}\right)  \label{eq:qo}
 \end{eqnarray}
$\mathcal{H}_{TW}$ is minimal for a non-zero number of opening base-pairs which increases as the stress strength increases (see figure \ref{fig:qeff}).\\
The total effective Hamiltonian is given by $\mathcal{H}_{eff}(\{\theta_i\})=\mathcal{H}_{ZB}(\{\theta_i\}) +\mathcal{H}_{TW}(n_o)$. We fix $\Delta g_{loop}=20 k_b T$.

\subsubsection{Torque-imposed ensemble}

The Benham Hamiltonian is defined in a superhelical density ($\sigma$)-imposed ensemble defined by equation \ref{eq:she}. In this section, we briefly discuss similar model but in the torque-imposed ensemble. \\
In this ensemble, a linear DNA segment (length $N$) is constraint by a weak torque $\Gamma$ applied on base $N$, the first bp being fixed. For each bp $i$, we define $\omega_i$ its orientation in the plane perpendicular to the average axis of the double-helix and oriented in the 5' to 3' direction (for a denatured bp-step, $\delta \omega_i\equiv\omega_{i+1}-\omega_i=\tau_i/(2\pi)$ of the Benham model). The total Hamiltonian of the system is then given by \cite{orland}
\begin{eqnarray}
\mathcal{H}&=&H_{ZB}+\frac{C}{2}\sum_i (1-\theta_i)(\delta\omega_i)^2 \nonumber \\
& &+\frac{K_r}{2}\sum_i \theta_i (\delta\omega_i-\omega_0)^2-\Gamma(\omega_N-\omega_1)
\end{eqnarray}
with $K_r=K'/(4\pi^2)$ and $\omega_0=2\pi/A$ the natural helical twist. 
Writing $(\omega_N-\omega_1)=\sum_i \delta\omega_i$, and integrating over the $\delta\omega_i$, leads to the effective Hamiltonian (relatively to the situation without torque):
\begin{equation}
\mathcal{H}_{eff}=\mathcal{H}_{ZB}+\left[\Gamma^2\left(\frac{1}{2C}-\frac{1}{2K_r}\right)-\Gamma \omega_0\right]\sum_i \theta_i
\end{equation}
Applying the constant torque $\Gamma$ is therefore equivalent to adding an external field $h_e=-[\Gamma^2(1/(2C)-1/(2K_r))-\Gamma \omega_0]$ in the ZB formalism. 

\subsection{Self-consistent linearization}

\subsubsection{Solving the Benham model}

Recently, Jeon et al \cite{jeonPRL10} derived formulas to solve the Benham model for homogeneous and random heterogeneous sequences, including the computation of sequence-average bubble properties. The computation is based on a reorganization of the partition function sum into partial sums for fixed numbers of bubbles. 

For heterogeneous sequences, Fye and Benham \cite{fye99} proposed an exact $\mathcal{O}(N^2)$-algorithm by decomposing the prefactor $\exp[-\beta\mathcal{H}_{TW}(n_o)]$ in discrete Fourier modes, and by using the transfer-matrix method. 
Benham and coworkers \cite{benham04,bi03,bi04} have also developed an approximate method which first involves a windowing procedure to find the minimum free energy and then consider only the states whose energies do not exceed the minimum one by more than a given threshold. At high threshold values or high negative superhelicity, the computation time for this algorithm scales exponentially with the threshold and the superhelical stress.
Both schemes are still time demanding for very long sequences.

\subsubsection{Self-consistent field}

 In order to speed up the resolution of the Benham model and to access directly to position-dependent opening properties of bps {\it and} bubbles, we develop an efficient variational method \cite{vologodskii} allowing us to use the transfer-matrix solution of the ZB model. For long sequences, assuming that fluctuations of $n_o$ are small around the value $\bar{n}_o$, we can expand $\mathcal{H}_{TW}(n_o)$ around $\bar{n}_o$. The approximated effective Hamiltonian then takes the typical ZB form:
\begin{equation}
 \mathcal{H}_{eff}\approx \mathcal{H}_{ZB}(\{\theta_i\})+\mathcal{H}_{TW}(\bar{n}_o)+(N-\bar{n}_o)h -h\sum_{i=1}^N \theta_i  \label{eq:heff}
 \end{equation}
 where $h=(\partial \mathcal{H}_{TW}/\partial n_o)(\bar{n}_o)$ represents the mean-field of our approximation. If one imposes the superhelical stress $\Gamma$ (ie, the torque) instead of the superhelical density, $h$ is related to the effective field $(4\pi^2N/K-1/C)(\Gamma^2/2)+2\pi\Gamma/A$ generated by the torque (see above). \\
In the following, we employ the Benham ensemble of an imposed superhelical density, $\sigma$, and determine $h$ (or $\Gamma$) self-consistently. Self-consistency requires $\langle n_o \rangle(\bar{n}_o)=\bar{n}_o$, and is equivalent to solving 
\begin{equation}\label{eq:fh}
h=(\partial \mathcal{H}_{TW}/\partial n_o)(\langle n_o \rangle(h)),
\end{equation}
 because the function $(\partial \mathcal{H}_{TW}/\partial n_o)(n_o)$ is monotonic.

\subsubsection{Homogeneous sequences}

\begin{figure}
\begin{center}
\includegraphics[width=8.5cm]{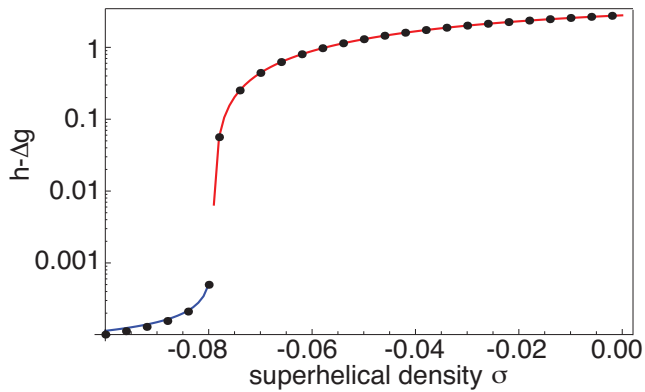}\caption{Numerical (dots) or analytical (lines) solutions of the self-consistency equation Eq.\ref{eq:fh} for a homogeneous sequence ($\Delta g=-3.14 k_B T$).}\label{fig:approx}
\end{center}
\end{figure}

For homogeneous sequences, the general solution of the self-consistency equation \ref{eq:fh} cannot be computed analytically. However, at low temperatures, weak superhelical densities and in the limit of infinitely long chains, a small perturbation development is valid and leads to analytical expressions for $h$. 
For infinitely long chains, $\mathcal{H}_{TW}$ becomes
\begin{equation}
\mathcal{H}_{TW}(x)/N=\frac{2 C K' \pi^2 (\sigma+x)^2}{A^2(4C \pi^2+K' x)}-f\,x
\end{equation}
with $x=n_o/N$, $K'=N K$ and $f=-\log[2\pi/(\beta C)]/(2\beta)$.
Assuming that the fraction of open base-pairs $x$ is small ($x<<1$), $\partial \mathcal{H}_{TW}/\partial n_o=\partial (\mathcal{H}_{TW}/N)/\partial x$ is given by
\begin{equation}\label{eq:dh}
\left(-f+\frac{K' \sigma}{A^2}-\frac{K'^2\sigma^2}{8 A^2 C \pi^2}\right)+K'\left(\frac{[K'\sigma-4\pi^2 C]^2}{A^2 (4\pi^2 C)^2}\right)x+o(x^2)
\end{equation}
Using Eq.\ref{eq:homozb} and noting $y=\exp[\beta h]$, we also have
\begin{equation}
x\approx \frac{s y }{(g y -1)^2}
\end{equation}
In the limit $g y >>1$, inserting $x\approx s/(g^2 y)$ in Eq.\ref{eq:dh} and solving Eq.\ref{eq:fh}, leads to
\begin{equation}\label{eq:h1}
h^* = -f+\frac{K'\sigma (8C \pi^2-K'\sigma)}{8A^2 C \pi^2}
\end{equation}
This expression is valid until $g y \sim 1 $ ($h\approx \Delta g$), i.e.
\begin{equation} 
\sigma\approx \left(\frac{2\pi}{K'}\right)\left(2\pi C-\sqrt{2C(2\pi^2 C-A^2[f-\Delta g])}\right)\equiv \sigma_l
\end{equation}
For $\sigma < \sigma_l$, we could write $g y =1+\epsilon $ (with $\epsilon >0 $). Then, $x\approx (s/g)/\epsilon^2$ and $h\approx \Delta g +\epsilon$. Solution of Eq.\ref{eq:fh} leads to
\begin{widetext}
\begin{equation}\label{eq:h2}
h^* = \Delta g+\left(\sqrt{\frac{K' s}{2\pi^2 C g}}\right)(4 C \pi^2-K'\sigma) \left(\frac{4 C \pi^2-K'\sigma}{\sqrt{8 A^2 \pi^2 f C+K'\sigma(K'\sigma-8\pi^2 C)+8A^2 \pi^2 C \Delta g}}\right)
\end{equation}
\end{widetext}
Figure \ref{fig:approx} shows that the numerical solution of Eq.\ref{eq:fh} agrees very well with the two expressions found above (Eq.\ref{eq:h1} and \ref{eq:h2}). \\
To compute bp or bubbles properties, formulas \ref{eq:homoz}, \ref{eq:homozb} and \ref{eq:homop} are still available if one replaces $g$ by $g\,y$ and $s$ by $s\, y$.

\subsubsection{Heterogeneous sequences}

For heterogeneous sequences, we use the bisection method coupled to the Newton-Raphson method \cite{press} to numerically solve the self-consistency equation, for fixed values of the temperature and of the superhelical density. Knowing that an evaluation of the function $f$ requires one transfer matrix method computation ($\mathcal{O}(N)$ each), it takes typically $10-20$ evaluations  to determine the root with a relative precision of $10^{-4}$. This allows numerically efficient computation of denaturation profiles. For example, computing the local closing probabilities for the {\it E.coli} genome ($N\sim4.6$ Mbps) takes about $70$ seconds on a 2.4 GHz computer with the self-consistent method, whatever the density is. On the same computer, it would take about $10^{10}$ s with the exact method ($\mathcal{O}(N^2)$ algorithm) for any $\sigma$ values (interpolation of data given in Ref.\cite{fye99}), and about $6.10^4$ s with the approximate method for $\sigma=-0.055$ ( and around 40 times more for $\sigma=-0.075$) \cite{benham04,bi03,bi04}. 
\begin{figure}
\begin{center}
\includegraphics[width=8.5cm]{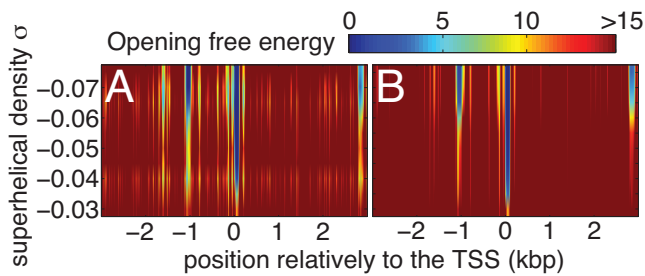}\caption{(Color online) Opening free energy $-\log[1-\langle \theta_i\rangle]$ for the extended neighborhood of the TSS of Fig. \ref{fig:land}, computed using the Benham's web server \cite{benhamweb} (A) or using our formalism (B).}\label{fig:compa}
\end{center}
\end{figure}

\subsubsection{Comparison with the Benham model}

The self-consistent linearization consists in working in the {\em torque}-imposed ensemble and determining the torque self-consistently to better approximate the {\em superhelical density}-imposed ensemble. 
This representation has the advantage of decoupling the opening of different parts of the molecule and is probably the simplest way to express the destabilizing effect of undertwisting on DNA stability.
In the thermodynamic limit, i.e., for very long (genomic) sequences and at low temperature (well below the melting temperature), we expect our linearization to be a quasi-exact solution of the Benham model due to the asymptotic decrease of fluctuations within the system. 
For shorter sequences, however, the non-linearity of the effective superhelical Hamiltonian $\mathcal{H}_{TW}$ (see Eq.\ref{eq:qo}) may significantly couple remote domains along the sequence, leading for example, to the closing of an open domain as one increases the superhelical stress (as observed in Fig.\ref{fig:compa}A). Although, the self-consistent linearization neglects such effects, it gives a reliable general picture of the superhelically-stressed destabilization of DNA sequences. In figure \ref{fig:compa}, we show a comparison of results obtained from the Benham web server \cite{bi04,benhamweb} and by our method. The agreement is excellent and the small deviations are mostly due to the slightly different parametrizations of the ZB parts and to the absence of finite size effect in our approach as discussed above.

\section{Results and Discussion}

In the following, we discuss results obtained for random homogeneous and heterogeneous sequences, as well as for some bacterial genomes ({\it E.coli}, {\it T. whipplei}, {\it A. Baumanii}, {\it B. subtilis} and {\it S. coelicolor}, see Table \ref{tab:zs}).
The results shown for random heterogeneous sequences were obtained by compiling profiles from 100 random sequences each containing $10^6$ bp. 

\subsection{Melting of superhelical DNA}

The melting properties of constrained DNAs reflect a balance between the two parts of $\mathcal{H}_{eff}$. At $37^o$ C, $\mathcal{H}_{ZB}$ opposes local opening the more strongly the higher the  GC-content of the sequence. In contrast, $\mathcal{H}_{TW}$ is minimal for a finite number of open base-pairs, which increases as $\sigma$ becomes more negative. At biological levels, the free energy of completely closed DNA becomes prohibitively large (see Fig.\ref{fig:qeff}). As a consequence, superhelicity leads to a notable level of base-pair opening at physiological temperatures, where unconstrained DNA exhibits negligible breathing. 

\begin{figure}
\begin{center}
\includegraphics[width=8.5cm]{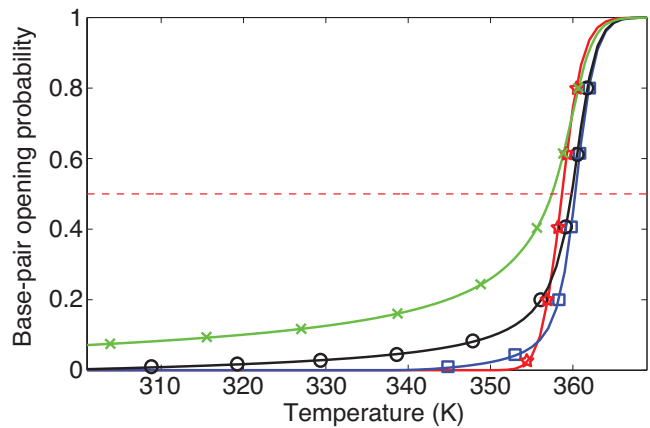}\caption{Evolution of the global opening probability $1-\Theta$ as a function of temperature for an unconstrained (stars) or constrained ($\sigma=0$: squares, $\sigma=-0.06$: circles, $\sigma=-0.2$: crosses) random sequence (GC=0.5).}\label{fig:thetatm}
\end{center}
\end{figure}
Figure \ref{fig:thetatm} shows the overall impact on the opening probability of imposing a superhelical constraint. For comparison, we have also included a melting curve for unconstrained DNA. At physiological temperatures, the unconstrained DNA is very stable whereas the superhelicity significantly contributes to opening of base pairs and bubbles. However, for intermediate and weak negative stresses,  the destabilizing effect of an imposed superhelical density is reversed close to the melting temperature due to the overall stabilizing impact of untwisting on the rest of the DNA ($\mathcal{H}_{TW}(n_o=N/2)>0$), resulting in a slowdown of the melting process via a change of $\Gamma$ (or $h$) with temperature. This effect has already been pointed out for positively-stressed homogeneous molecules \cite{benham96}. For stronger stresses, the effective twisting Hamiltonian $\mathcal{H}_{TW}$ at the melting transition ($n_o=N/2$) is still negative and then results in a decrease of the melting temperature. A quantification on this salt-, GC-dependent effect is described below.

\begin{figure}
\begin{center}
\includegraphics[width=8.5cm]{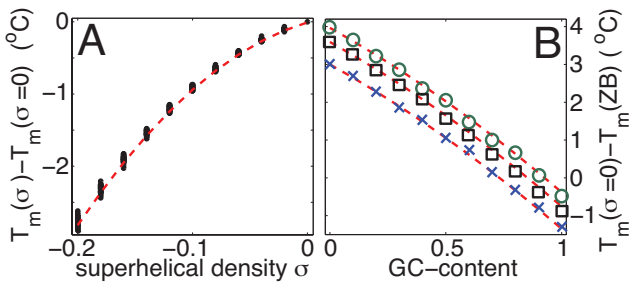}\caption{(A) Melting temperature for random sequences (GC$\in[0,1]$) at different salt concentrations ($[Na^+]\in[0.05,1]$ M) as a function of the superhelical density $\sigma$ (dots), respectively to the situation with $\sigma=0$. (B) Melting temperature at $\sigma=0$ for random sequences as a function of their GC-content, at $[Na^+]=0.05$ (circles), 0.1 (squares) and 0.3 M (crosses), respectively to the corresponding unconstrained (ZB) situation. Dashed lines represent the fitted relation given in Eq.\ref{eq:tmsiggc}.}\label{fig:salt}
\end{center}
\end{figure}
On figure \ref{fig:salt} A, we observe that, under the different considered GC-contents and salt concentrations, the melting temperature of random superhelical DNA is a quadratic function of $\sigma$, and only the intercept of this function ($T_m(\sigma=0)$) depends on the GC and on $[Na^+]$ (see figure \ref{fig:salt} B). From a systematic study of the melting temperature $T_m$ as a function of the GC-content and the salt concentration, we fitted the empirical relation (in $^o$ C):
\begin{eqnarray}\label{eq:tmsiggc}
\Delta T_m([Na^+],\textrm{GC},\sigma) & = & 2.35 -0.54\log[Na^+]-3.42\textrm{GC} \nonumber \\
& & -0.93\textrm{GC}^2+4.7\sigma-46.8\sigma^2
\end{eqnarray}
where $\Delta T_m$ is the difference in melting temperature between a constrained and an unconstrained random DNA polymers.  

\begin{figure}
\begin{center}
\includegraphics[width=8.5cm]{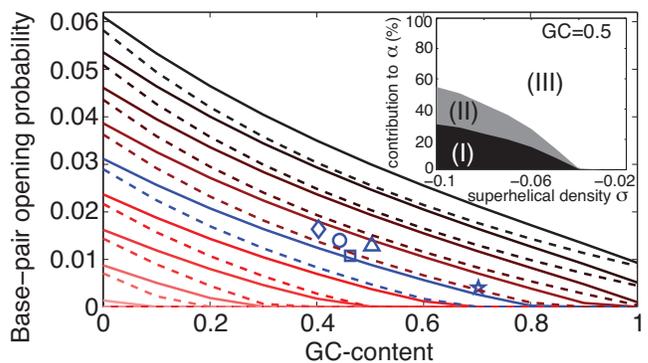}\caption{(Color online) Evolution at $T=37^o$ C of the global opening probability $1-\Theta$ for superhelically constrained random heterogeneous (full lines) and homogeneous (dashed lines) DNAs under different $\sigma$ (from -0.02 to -0.1 every 0.01: color scale from light to dark red, $\sigma=-0.06$: blue line) and for several bacterial genomes at $\sigma=-0.06$: {\it E. coli} (triangles), {\it T. whipplei} (squares), {\it A. Baumanii} (diamonds), {\it B. subtilis} (circles) and {\it S. coelicolor} (stars). Inset: proportion of the different contributions to the linking number difference $\alpha$: relaxation due to denaturation $-n_o/N$ (I), over-twist of the denatured regions $\mathcal{T}$ (II) and residual linking number $\alpha_r$ (III) (see Eq.\ref{eq:she}).}\label{fig:stresrandt}
\end{center}
\end{figure}
Figure \ref{fig:stresrandt} shows the dependance of the opening probability $1-\Theta$ as a function of the GC-content. Results are reported  for different levels of superhelical density and at $T=37^o$ C, in the biological relevant range, i.e the overall degree of opening is small yet strongly increased relative to the unconstrained case (see also Fig.\ref{fig:thetatm}). We have included results for both homogeneous and heterogeneous systems. In the former case, the employed NN-free energies $\Delta g$ were determined as composition dependent averages over the tabulated step parameters \cite{SL,jostBJ09}.\\
Passing a threshold \cite{jeonPRL10} (see Inset in Fig.\ref{fig:stresrandt}) depending on the GC-content, strong $\sigma$-values allow the opening of many bp along the sequences. 
For a fixed superhelical stress, as expected, $1-\Theta$ is a decreasing function of the GC-content. 
Differences between homogeneous and heterogeneous are weak, meaning that sequence heterogeneity self-averages and has only a small effect on the total degree of opening.\\
The inset in figure \ref{fig:stresrandt} shows the different contributions to the linking number difference as a function of the superhelical density (see Eq.\ref{eq:she}) for random sequences with GC=0.5. The over-twist $\mathcal{T}$ contribution is estimated by integrating over the $n_o$ continuous degrees of freedom $\tau_i$ \cite{fye99} and applying the self-consistent linearization: 
\begin{eqnarray}
\langle \mathcal{T} \rangle & = &\sum_{\{\theta_i \}} \frac{K n_o}{4\pi^2 C+K n_o}(\alpha+n_o/A) \frac{e^{-H_{eff}(\{\theta_i\})}}{\mathcal{Z}}\\
& \approx & \frac{K \langle n_o \rangle }{4\pi^2 C+K \langle n_o \rangle}(\alpha+\langle n_o \rangle/A) 
\end{eqnarray}
We observe that the over-twist increases linearly with the opening probability. While for weak stresses, almost all the superhelical energy is stored in the (residual) deformations of the double-helix, for strong stresses, more than 50 \% of the imposed superhelical constraint is used to drive the local denaturation of bps. Accounting for the writhe in the model should however decrease the contributions due to bp-denaturation and over-twisting since a part of the constraint would be absorbed by coils of the double-helix.

\subsection{Sequence heterogeneity and bubble statistics in superhelical DNA}

\subsubsection{Bubble statistics in unconstrained DNA}

\begin{figure}
\begin{center}
\includegraphics[width=8.5cm]{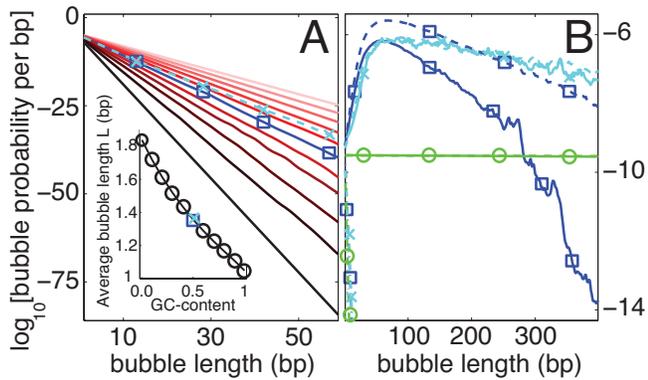}\caption{(Color online) Evolution of the bubble probability per bp $P_l$ to observe a bubble of length $l$. (A) For unconstrained molecules: random sequences of different GC-content (GC from 0 to 1 every 0.1: color scale from light to dark red, GC=0.5: squares) and the genome of {\it E. coli} (dashed line with crosses, GC=0.51). Inset: average bubble length $\mathcal{L}$ as a function of the GC-content. (B) For superhelically-stressed molecules: a random sequence ($GC=0.5$, squares), the genome of {\it E. coli} (crosses) and a homogeneous sequence (with $\Delta g=-1.9 k_B T$ equals to the average NN-free energies in {\it E. coli}, circles), and $\sigma=0$ (dotted and dashed lines), $-0.06$ (full lines) and -0.1 (dashed lines). }\label{fig:probzb}
\end{center}
\end{figure}
Figure \ref{fig:probzb}A shows the bubble probabilities computed with the ZB model for random DNAs of different GC-content, with $\Delta g_{loop}=10k_bT$ consistent with very small loops ($\bar{L}\sim \mathcal{L}\sim 1$, see inset in Fig.\ref{fig:probzb} A). We remark an exponential decrease of $P_l$ with very short decay lengths, corresponding to fairly closed molecules. Increasing GC-content stabilizes the DNA and reduces average bubble lengths. Even accounting for the weak biological sequence effect apparent in our results, the absolute level of bubble opening in unconstrained DNA seems too small for natural breathing to play a direct biological role \cite{benham06}.

Whereas we obtain similar $\mathcal{L}$-values as reported for the Peyrard-Bishop-Dauxois (PBD) model \cite{ares07}, absolute probabilities $P_l$ are far lower (for example $P_l\sim 10^{-11}$ (ZB) versus $\sim 10^{-5}$ (PBD) for $l=10$ and GC=0.5), mainly due to the absence of a cooperativity penalty factor preventing bubble formation in the PBD model.

\subsubsection{Bubble statistics in superhelical DNA}

Figure \ref{fig:probzb}B highlights the huge impact of the superhelical stress on the bubble statistics. For physiological levels, the opening probability per bp remains significant even for large bubbles and has increased by many orders of magnitude compared to the unconstrained situation. For example, for GC=0.5 and $\sigma=-0.06$, $P_{l=10}\approx 10^{-9}$ and $P_{l=60}\approx 10^{-7}$ for superhelical DNA, while for the corresponding unconstrained DNA, we found $P_{l=10}\approx 10^{-11}$ and $P_{l=60}\approx 10^{-41}$.

\begin{figure}
\begin{center}
\includegraphics[width=8.5cm]{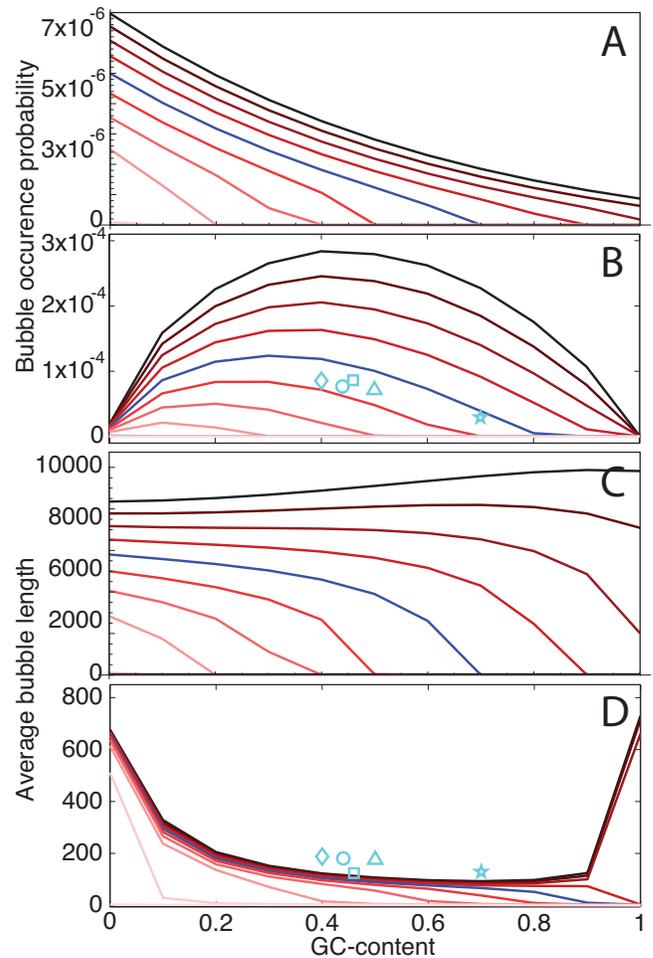}\caption{(Color online) Evolution of bubble occurrence probability $P_b$ (A,B) and of the average bubble length $\mathcal{L}$ (C,D) for superhelically constrained random heterogeneous (B,D) and homogeneous (A,C) DNAs under different stresses $\sigma$ (from -0.02 to -0.1 every 0.01: color scale from light to dark red, $\sigma=-0.06$: blue line) and for several bacterial genomes at $\sigma=-0.06$: {\it E. coli} (triangles), {\it T. whipplei} (squares), {\it A. Baumanii} (diamonds), {\it B. subtilis} (circles) and {\it S. coelicolor} (stars).}\label{fig:stresrand}
\end{center}
\end{figure}
\begin{figure}
\begin{center}
\includegraphics[width=8.5cm]{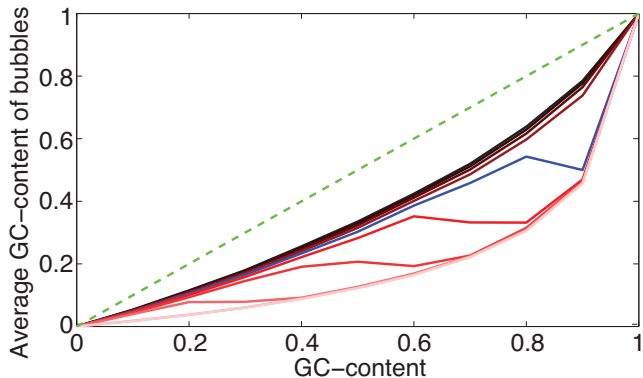}\caption{(Color online) Evolution of the average GC-content of bubbles for superhelically constrained random heterogeneous DNAs under different stresses $\sigma$ (from 0 to -0.1 every 0.01: color scale from light to dark red, $\sigma=-0.06$: blue line). The dashed line represents results for homogeneous sequences.}\label{fig:gcloc}
\end{center}
\end{figure}

Figure \ref{fig:stresrand} shows the evolution of the bubble occurence $P_b$ and of the average bubble length $\mathcal{L}$ for different $\sigma$-values and GC-content.
With $\mathcal{H}_{TW}(n_0)$ only being a function of the total number of open base-pairs, bubble sizes in homogeneous systems are determined by a competition between the bubble initiation penalty, $\Delta g_{loop}$, favoring the opening of a small number of large bubbles, and entropy, favoring the opening of a large number of small bubbles. In heterogeneous systems, it is possible to lower the fraction of stable GC-steps in the open domains by denaturing a larger number of smaller bubbles in particularly AT-rich regions (see Fig.\ref{fig:gcloc}) \cite{hwa03}. The comparison in Figs. \ref{fig:stresrand} and \ref{fig:gcloc} shows that the disorder effect dominates in the present case \cite{garel05,coluzzi07,monthus07} with the number of bubbles being maximal around GC=0.5.  For biological superhelicities, $\mathcal{L}\approx700$ for random AT- and GC-DNAs, while $\mathcal{L}\approx150$ for $0.2<f_{GC}<0.9$.

Sequence-heterogeneity plays an essential role by lowering the fraction of stable GC-steps in the open domains and leads to a {\em localization} of open base-pairs in (AT-enriched) stress-induced duplex destabilized regions \cite{wang04,ak05,wang08}, whose length in turn limits the bubble sizes. 
Indeed, figure \ref{fig:gcloc} shows that bubbles appear mainly in AT-enriched regions compared to the background GC-content. Interestingly, for intermediate (biological) $\sigma$ values, the evolution of the GC-composition of bubbles remains flat over a large range of global GC-content. 

\subsection{Bubble statistics in biological DNA}

Compared to random sequences with identical GC-content, bacterial genomes ({\it E.coli}, {\it T. whipplei}, {\it A. Baumanii}, {\it B. subtilis} and {\it S. coelicolor}) exhibit higher averages degrees of opening and increased average bubble sizes (symbols in Figs.\ref{fig:stresrandt} and \ref{fig:stresrand}).
Figure \ref{fig:probzb}B shows striking discrepancies in the bubble distribution $P_l$ between random and biological sequences,  with a significant enrichment of large bubbles. In the latter case, the nearly flat distributions resemble those expected for homogeneous sequences and can be viewed as a signature of large stress-induced destabilized domains, which concentrate the DNA breathing into large regions with homogeneous opening profiles. We also note that the level of opening in the biological domains is comparatively insensitive to the precise level of the superhelical density.

\begin{table}
\caption{Z-scores computed relatively to the average bubble length for 5 prokaryotic organisms.}\label{tab:zs}
\begin{tabular}{l|ccc}
\hline
Organism & $N$ (Mbp) & GC-content & Z-score \\
\hline
{\it A. baumanii} & 3.98 & 0.40 & 23.0 \\
{\it B. subtilis} & 4.21 & 0.44 & 21.0 \\
{\it E. coli} & 4.64 & 0.50 & 25.7 \\
{\it S. coelicolor} & 8.67 & 0.70 & 17.6\\
{\it T. whipplei} & 0.93 & 0.46 & 10.6 \\
\hline
\end{tabular}
\end{table}
In general, biological sequences are more destabilized than random sequences with a same GC-content, leading to less but longer bubbles. We estimate these differences by computing, for all genomes, the Z-score relatively to $\mathcal{L}$ (see Tab.\ref{tab:zs}). The Z-score is an estimation of the non-randomness of a specific sample. For the average bubble length $\mathcal{L}$, it can be computed by
\begin{equation}
Z=\frac{\mathcal{L}(\mathrm{genome})-\langle\mathcal{L}(\mathrm{random})\rangle}{\sigma_{\mathcal{L}(\mathrm{random})}}
\end{equation}
where $\langle\mathcal{L}(\mathrm{random})\rangle$ and $\sigma_{\mathcal{L}(\mathrm{random})}$ are the mean and the standard deviation of $\mathcal{L}$ computed on 100 random sequences with the same GC and length as the corresponding genome. The Z-score represents therefore the distance between a specific point and the mean value obtained for random sequences, given in standard deviation units.
Z-scores of 20 for $\mathcal{L}$ suggests the presence of over-represented long AT-rich domains in bacterial genomes which have the ability to easily open under superhelical stress.
Interestingly, the organism with the lowest Z-score ({\it T. whipplei}) was shown to exhibit a random-like behavior relatively to the local melting temperature distribution \cite{jostJPCM09}.

\subsection{Bubble positioning in the promoter regions of biological DNA}

\begin{figure}
\begin{center}
\includegraphics[width=8.5cm]{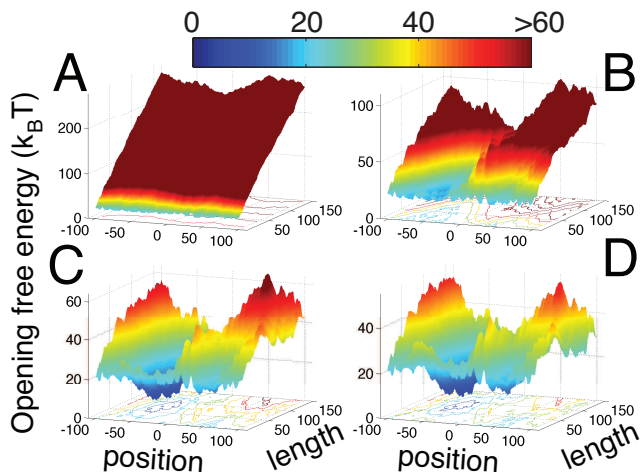}\caption{(Color online) Free energies to open a bubble of a given length centered around a given nucleotide ($-k_B T\log[P_{i,l}]$) for the neighborhood of a transcription start site (bp n$^o$ 259382) in the {\it E. coli} genome, in the presence (B, C, D) or in the absence (A) of an imposed superhelical density $\sigma=-0.03$ (B), -0.06 (C) and -0.10 (D).}\label{fig:land}
\end{center}
\end{figure}
Positions of strongly stress-induced destabilized regions have been shown to correlate with many regulatory regions \cite{wang08} including transcription start sites \cite {wang04} or origins of replication \cite{ak05}. In this section, we focus on bubble positioning inside the promoter regions of the bacterium {\it E. coli}.\\
The transcription of DNA is initiated by the local opening of the double-helix at transcription start sites. Figure \ref{fig:compa} illustrates their association with strongly stress-induced destabilized regions. In addition to position-dependent opening probabilities, our approach allows us to calculate the complete bubble free-energy landscape, $G_{i,l}=-k_BT \log P_{i,l}$. Figure \ref{fig:land} shows the effect of superhelicity on $G_{i,l}$ for the neighborhood of the same TSS in {\it E. coli}. The analysis reveals that opening is the result of the meta-stable unwinding of a large bubble and not of enhanced small scale breathing. We note that knowledge of $G_{i,l}$ is essential for modeling the dynamics of bubble nucleation and growth \cite{ambjornsson07}. 

\begin{figure}
\begin{center}
\includegraphics[width=8.5cm]{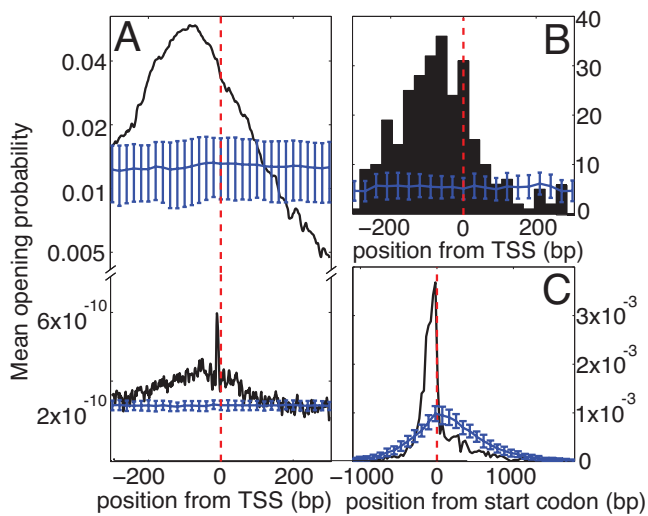}\caption{(A) Opening probability as a function of the position relatively to the TSS, averaged on 760 promoter regions of {\it E. coli} \cite{regulondb} (black line), or on 760 randomly picked regions inside the genome (blue line), for a constrained (top) or unconstrained (down) DNA. (B) Histogram of the highly probable ($P_{i,l} \ge 10^{-5}$) bubble centers included in the regions [TSS-300,TSS+300] for the 760 studied TSS (black bars) or for randomly picked regions (blue line). (C) Probability distribution function of the distance between highly probable bubble centers and the nearest start codons for the 1158 actually found (black line) or randomly situated (blue line) bubbles.}\label{fig:tss}
\end{center}
\end{figure}
Figure \ref{fig:tss} analyses the statistical relation beween TSSs and bubbles induced by superhelical stress for the entire genome of {\it E. coli}.
Figure \ref{fig:tss}A shows a significant and non-random destabilization around TSSs, with a maximal opening around $-80$. The same computation using the ZB model shows insignificant and orders-of-magnitude smaller opening probabilities. However, we find a non-random signal around TSS-10 corresponding to the position of the AT-rich Pribnow box, an essential motif to start transcription in bacteria \cite{pribnowPNAS75}.
Figure \ref{fig:tss}B gives the relative positions of highly probable bubbles included in TSS neighborhoods. 
The centers of these bubbles are mainly localized in the [TSS-200,TSS] region where many transcriptional and promoter factors are recruited and bind to DNA \cite{regulondb}. 
Conversely, figure \ref{fig:tss}C confirms that the majority of highly probable bubbles are situated upstream and close to start codons of genes. Actually, these bubbles are composed by around $36\%$ of coding bps, significantly lower than the percentage of coding bps in {\it E. coli} ($88\%$).

\section{Summary and Conclusion} 
We have developed a numerically efficient, self-consistent solution of the Benham model of bubble opening in superhelically constrained DNA. In particular, we are able to go beyond the calculation of opening probabilities for base pairs and to address the full, position-dependent bubble statistics for entire genomes. Our results indicate, that negative supercoiling leads to (meta-) stable unwinding of bubbles comprising ${\cal O}(100-1000)$ base-pairs. In heterogeneous sequences, bubbles are strongly localized in AT-rich domains with sequence disorder dominating the bubble statistics. As we have shown, large bubbles open with a significantly larger probability in biological sequences, than in random sequences with identical GC-content. In the case of {\it E. coli}, the most likely bubbles are located directly upstream from transcription start sites, highlighting the biological importance of this now well understood, physical property of DNA.


\end{document}